\documentclass[10pt,journal,compsoc,letterpaper]{IEEEtran}

\ifCLASSOPTIONcompsoc
  \usepackage[nocompress]{cite}
\else
  \usepackage{cite}
\fi

\usepackage{amsmath,amssymb,amsfonts}
\usepackage{algorithmic}
\usepackage{graphicx}
\usepackage{textcomp}
\usepackage{xcolor}
\usepackage{amsmath}
\usepackage{array}
\usepackage{multirow}
\usepackage{color}
\usepackage{enumitem}
\usepackage{fancyhdr}
\usepackage{soul}
\soulregister\ref7
\soulregister\cite7
\soulregister\FW7
\soulregister\FPR7
\soulregister\FR7
\soulregister\AF7
\soulregister\footnote7
\usepackage[normalem]{ulem}

\setlist[itemize]{leftmargin=*}

\setitemize{noitemsep,topsep=6pt,parsep=6pt,partopsep=0pt}

\usepackage{tikz}

\newcommand*\mybtextcircled[1]{\tikz[baseline=(char.base)]{\node[shape=circle,fill,inner sep=1pt] (char) {\small \textcolor{white}{#1}};}}

\usepackage[moderate,leading=normal,tracking=normal,wordspacing=normal,lists=normal,bibnotes=normal,mathspacing=tight]{savetrees}

\newcommand{\UF}{Unfused} 
\newcommand{\FW}{FusedWrite} 
\newcommand{\FPR}{FusedPartialReduction} 
\newcommand{\FR}{FusedOCG} 
\newcommand{\AF}{FusedIOCG} 

\begin{document}

\title{Making Convolutions Resilient via Algorithm-Based Error Detection Techniques}

\author{Siva~Kumar~Sastry~Hari, 
	Michael~B.~Sullivan, Timothy~Tsai, and Stephen~W.~Keckler \\ NVIDIA Corporation
}

\IEEEtitleabstractindextext{%
\begin{abstract}

The ability of Convolutional Neural Networks (CNNs) to accurately
process real-time telemetry has boosted their use in safety-critical
and high-performance computing systems. As such systems
  require high levels of resilience to errors, CNNs must execute
  correctly in the presence of hardware faults.  Full duplication
provides the needed assurance but incurs a prohibitive 100\% overhead.
Algorithmic techniques are known to offer low-cost solutions, but the
practical feasibility and performance of such techniques have never
been studied for CNN deployment platforms (e.g., TensorFlow or
TensorRT on GPUs). In this paper, we focus on algorithmically
verifying Convolutions, which are the most resource-demanding
operations in CNNs. We use checksums to verify convolutions, adding a
small amount of redundancy, far less than full-duplication. We first
identify the challenges that arise in employing Algorithm-Based Error
Detection (ABED) for Convolutions in optimized inference platforms
that fuse multiple network layers and use reduced-precision
operations, and demonstrate how to overcome them.  We propose and
evaluate variations of ABED techniques that offer implementation
complexity, runtime overhead, and coverage trade-offs.  Results show
that ABED can detect all transient hardware errors that might
otherwise corrupt output and does so while incurring low runtime
overheads (6-23\%), offering at least 1.6$\times$ throughput to
workloads compared to full duplication.

\end{abstract}

}

\maketitle

\thispagestyle{empty}

\IEEEraisesectionheading{\section{Introduction}\label{sec:introduction}}

Following recent improvement in the ability of Convolutional Neural Networks
(CNNs) to perform complex tasks with high efficiency and accuracy, CNNs have
made their way into safety-critical and High Performance Computing
(HPC) systems. For example, autonomous vehicles (AVs) employ CNNs to perform
complex tasks such as vehicle, cyclist, 
pedestrian, lane, road-sign, and
free-space detection
\cite{apollobaidu, driveav}.  HPC systems also employ CNNs for 
object classification and detection, image segmentation, and
video analytics for application domains such as healthcare, climate analysis,
and surveillance, and often in real-time settings~\cite{TeslaT4,
kurth2018exascale}.  As the compute throughput and power efficiency demands of
the CNN-based safety-critical and HPC systems are high, efficient platforms are
being designed to meet the throughput demands within limited power
budgets~\cite{Lin2018, Sze2018}.  For example, the recently released NVIDIA DRIVE
AGX Xavier System-on-Chip and T4 GPU deliver up to 32 and 130 TOPS while
consuming just 30 and 70 watts of power, respectively~\cite{driveav, TeslaT4}.

Safety-critical and HPC systems must be designed to tolerate hardware errors
such as those originating from hardware transient, intermittent, and permanent
faults. Some market segments require systems to meet strict safety standards,
such as the ISO 26262 functional safety standard for AVs~\cite{ISO26262}. This
standard requires the system to be robust to single-point transient,
intermittent, and permanent faults either by design or by coverage from safety
procedures (such as ECC and parity). The level of robustness a hardware
component desires to obtain is determined by
the Automotive Safety Integrity Level (ASIL). For ASIL D (the highest safety level), the system is required to be
robust to $\ge$99\% of faults; the requirements for ASIL C and B are 97\% and
90\%, respectively~\cite{Nardi2017}.  The rate of residual failures, measured
in Failure In Time or FITs (where 1 FIT refers to 1 failure per billion hours),
must also be $\le$100 and $\le$10 FIT for ASIL B/C and D, respectively.  While
the HPC market also demands high resilience, the requirements are not as
rigorous as those of ISO 26262~\cite{Snir2014}. 

With the increasing prevalence of CNNs in safety-critical and HPC systems and
the resilience requirements of such systems, correctness of CNNs must be
assured in the presence of hardware faults for safety and standards-compliance.
Prior studies have analyzed the effects of hardware errors on CNN outputs and
observed noticeable corruptions that must be mitigated to ensure safe
operation~\cite{Reagen2018, Li2017}. 
Processors deployed in such systems employ ECC and/or parity in major SRAM
structures. This protection is typically not sufficient to meet the
requirements of ASIL B, C, or D for all the hardware error sources. 
Aggressive employment of ECC/parity on flip-flops and small SRAM buffers comes at 
an area cost and may still be insufficient for all error sources due to the
error rate contribution from non-storage elements. For intermittent and
permanent faults, 
non-storage elements contribute significantly towards the total error
rates in  GPUs and DNN-accelerators that dedicate significant
chip area to logic~\cite{Constantinescu2008}.  Full hardware redundancy 
can provide the needed safety~\cite{Alcaide2019, TeslaFSDComputer, ARMCortexR5},
 but it reduces the throughput by
2$\times$ or more, which is prohibitive for resource-constrained systems.  The
goal of this paper is to develop a low-cost CNN-specific resilience solution
that allows the full system to meet the target markets' requirements, while
incurring far lower overheads compared to full duplication.

Over 90\% of the computation during CNN inference and training is in
convolutions~\cite{Li1702}.
Algorithmic methods have been
devised to speed up the convolution operation. These methods include using
General Matrix Multiplication (GEMM), Fast Fourier Transformation (FFT), and
Winograd~\cite{Sze2018}. 
Fault tolerance approaches that leverage
algorithm knowledge, known as algorithm-based fault tolerance (ABFT), have been
shown to provide lower overheads for GEMM and FFT than full duplication
algorithms~\cite{
Huang1984, hui_optimized_2018}. These techniques
leverage the fact that these operations are linear and verify the correctness
using a checksum-based approach.  These techniques compute checksums for input
data, store them with the original data, perform the original and redundant
computation, verify outputs, and possibly correct errors. The number of extra
compute operations they introduce is a small fraction of the original
computation.  While prior ABFT implementations have achieved runtime overheads
of about 20\% for square matrices~\cite{ding_matrix_2011}, our analysis shows
that the overheads can be much higher ($>$50\%) for the non-square matrices
that are typically used in CNNs. The main sources of overheads include running
the larger GEMM, managing storage to store checksums and associated data
movement (especially if they are stored with the input data), and computing the
checksums online for the input and output matrices. 

While ABFT techniques aim to correct errors inline along with detection, the
correction capability is limited (e.g., only a single cell error in a row or
column can be corrected) and there is no evidence that it is sufficient for
real hardware errors.  Moreover, detecting an error to prevent silent data
corruption (SDC) is more important to safety-critical and HPC systems than the
ability to correct it inline. Upon error detection, a low-cost local recovery
mechanism can be invoked that either restores the system
state~\cite{chung2012cds} or reruns the operation on the same or a spare
resource. For rare locally-unrecoverable errors, a heavy-weight fallback
mechanism can be invoked (e.g., transition to a degraded mode of
operation~\cite{safetyreport2019}) to handle the detected error. 

Based on this observation, we focus on algorithm-based error detection (ABED)
techniques in this paper and avoid additional costs associated with inline
error tolerance.  Since convolution is also a linear operation, like GEMM, and
a similar checksum-based solution can be applied at the convolution level.  A
recent study explored using such a solution to detect errors in CNN
accelerators (via hardware modifications) with the goal of overclocking the
system~\cite{Marty2018} and detecting corruptions online. It proposed using
checksums for filters and input feature maps (two inputs to the convolution) to
verify the checksum of the output. 

CNN inference deployment platforms (e.g., TensorRT, TensorFlow, PyTorch) are
commonly used in safety-critical and HPC systems~\cite{driveav, TeslaT4}.
Employing an algorithmic resilience technique in such platforms for seamless
application across architectures (e.g., CPUs, GPUs, or accelerators) is
desirable. However, several feasibility-, performance-, and coverage-related
challenges remain.  (1) The increasing use of reduced-precision data types
(e.g., 8 and 4-bit integers) in CNNs introduces new challenges for
checksum-based error detection techniques.  For example, the checksum
calculations can overflow without careful design.  (2) Convolution, a linear
operation, is often fused with subsequent activation layers, which are
non-linear operations, to reduce data movement and improve
performance~\cite{TeslaT4}. Checksum-based techniques do not not apply to
the non-linear computations and separating the linear and non-linear
computations into different operations can incur high overheads, introducing
additional design challenge.  

We address overflow in checksum calculation and storage such that no
information is lost, and full error detection capability is maintained.  We
also present implementation options that enable ABED techniques to function
with the fused (linear and non-linear) layers.  We not only addressed the above
two challenges for the previously proposed ABED algorithm that employed
checksums for both filter and input feature map checksums~\cite{Marty2018}, but
also proposed and analyzed two other variations that use checksums just for
filters and input feature maps, respectively.  We leverage domain knowledge of
CNNs to simplify memory management to store checksums and reduce the number of
online tasks.  

We study the implementation complexity, runtime overhead, and resilience
related trade-offs offered by the three ABED techniques. For architecture that
protect the memory subsystem well with ECC/parity, a filter-only checksum-based
ABED technique offers a lower-overhead solution when used with commonly-used
CNN pruning optimization~\cite{Molchanov2016, Huang2018}. The filter and input
feature map checksum-based technique provides high coverage for architectures
that do not sufficiently protect memory subsystem with low overheads even when
deployed without CNN pruning optimization.  Our results show that ABED can
eliminate all transient-error-induced convolution output corruptions with low
(6-23\%) runtime overhead on state-of-the-art GPUs, offering at least
1.6$\times$ throughput improvements to workloads compared to full duplication.

\section{Background}
\label{sec:background}

\subsection{Related Work}

{\bf Full Hardware/Software Redundancy:} Traditional business-class systems
(e.g., IBM Z Series machines~\cite{Bartlett2004IBMZ}) employ expensive
hardware-managed dual- or triple-modular redundancy schemes. In safety-critical
systems, similar techniques are being employed to meet the highest-levels of
safety integrity requirements~\cite{ARMCortexR5, Alcaide2019, TeslaFSDComputer}\@.  
Software techniques have
been explored that introduce redundancy at different granularities including
the process, GPU kernel, thread, and assembly instruction
level~\cite{
Dimitrov2009, Wadden2014Redundancy,
SInRG2018}, but they all incur overheads that are high for
resource-limited real-time systems.  

\noindent
{\bf Algorithm-Based Fault Tolerance (ABFT):} ABFT techniques leverage
algorithmic knowledge to detect and correct errors with very low
overheads and they have been heavily studied in literature for GEMM
and FFT~\cite{Huang1984, hui_optimized_2018, ding_matrix_2011,
liang_correcting_2017}\@.
For
GEMM, these approaches introduce row checksums for one of the matrices
and column checksums for the other. When the matrix multiplication
operation is performed with the checksums, an output matrix is
produced with an extra row and column. Each of these extra values are
expected to match the checksums of the rows and columns of the output
matrix. In an event of an error, these techniques can localize the
error and use redundant information to correct certain types of
errors.  Several algorithmic techniques have been proposed to increase
the capability of correcting output matrix cell corruptions with
little emphasis on studying how low-level hardware faults manifest at
that level~\cite{Chen_2016, Wu_2016}\@. Hence it is unclear
whether the benefits of the additional correction capability outweigh
the costs of higher overhead.

A recent study showed that protecting GEMMs in CNNs via ABFT can provide high
protection~\cite{Santos2019}\@. While no runtime overhead analysis was included,
the paper acknowledged that GEMM kernels are tuned to fully
use caches and registers, and adding an extra dimension for checksum storage
would not only compromise execution time, but also significantly increase data
movement and memory latency.  Our experiments confirm this hypothesis and show that
the ABFT's overheads for non-square matrices commonly used in CNNs is high
($>$50\%).

Our results show that ABFT's error correction capabilities often introduce
higher overheads. A related study suggests that error detection and
re-computation can be cheaper than checksum-based ABFT
techniques~\cite{al-yamani_performance_2001}\@.  Moreover, the effectiveness of
the ABFT's correction capability is not established for real hardware errors
because they may not manifest as correctable (single-cell output) corruptions. 

Based on a similar observations, researchers proposed using a checksum-based
ABED technique~\cite{Marty2018} to detect errors during a convolution.  Since
convolution is a linear operation like GEMM, a checksum-based error detection
technique can be extended to it as well. This work used checksums for both the
filters and input feature maps to verify the output.  The goal of the paper was
to overclock a CNN accelerator and detect incorrectly computed convolutions. It
extended the hardware accelerator to include the detection technique.   
Several challenges arise while applying ABED to convolutions in
optimized CNN inference platforms (e.g., TensorRT, TensorFlow, PyTorch)
commonly used in safety-critical and HPC systems.  (1) The use of reduced
precision operations is common during inference and without proper care, the
checksum arithmetic can overflow.  (2) A convolution operation is often fused
with a non-linear activation layer, to reduce data movement and improve
performance~\cite{TeslaT4}\@. Checksum-based techniques apply only to the
linear operations. Separating the linear and non-linear computations into
different operations can incur high overheads due to additional data movement,
introducing additional implementation challenge.  (3) Without customizing ABED
to the CNN inference frameworks, online checksum storage management and
generation for the output and both the inputs on every convolution introduces
avoidable runtime overheads (explained further in Sections~\ref{method:abft}
and~\ref{results:abft}).  

We not only address these challenges in this paper, but also identified and
explored two other ABED variants (not previously studied) that use checksums
either for filters or input feature maps (explained in
Section~\ref{sec:approach}).  These variants offer interesting error coverage
and performance trade-offs, important in selecting an optimal solution for a
target safety-critical system. 

\begin{figure}[tbp]
 \centering
	\includegraphics[width=\columnwidth]{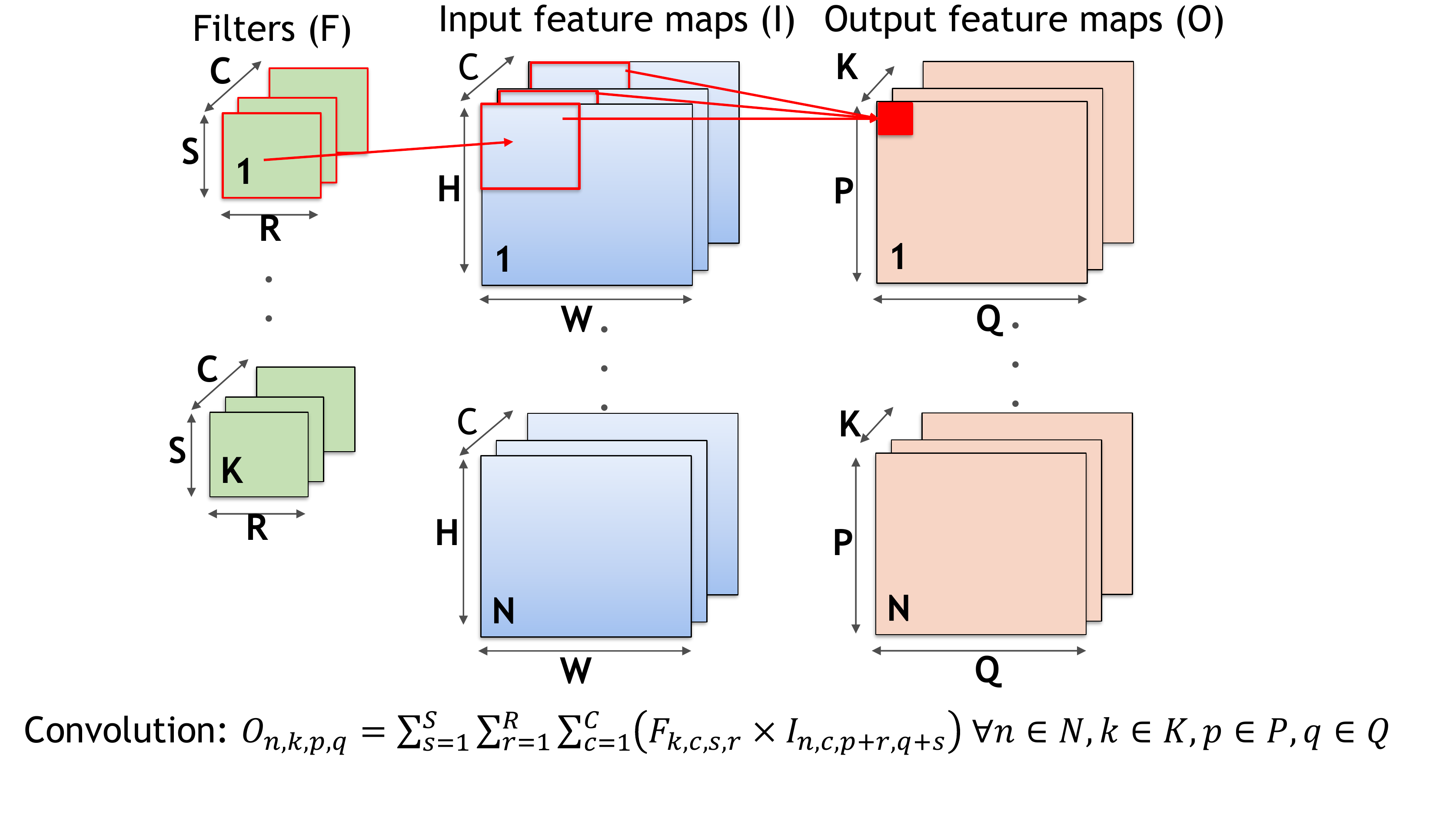}
	\vspace{-0.3in}
	\caption{A typical convolution operation used by most CNNs. }
	\label{fig:conv_background}
\end{figure}

\subsection{The Convolution Operation}

A convolution operation takes two input tensors and produces one output tensor.
One of the input tensors is for the input feature maps (or fmaps); these fmaps
are either the output of the previous layer or the input to the network.  Input
fmaps are represented as a 4-D tensor in most CNNs.  Each feature map is a 2-D
tensor with height (H) and width (W).  Typically, many feature maps are stacked
to form a 3-D tensor.  The number of channels (C) defines the number of feature
maps in the stack. Many 3-D input fmaps are batched (batch size = N) together
to form the 4-D feature map tensor.
The other input tensor is the set of filters, which consists of weights that are
computed during the training process. Each filter is a 3-D tensor of weights
with dimensions height (S), width (R), and channels (C). Each
convolution layer has multiple filters (number of filters = K), adding
an extra dimension to produce a 4-D tensor.

Each output feature map value is produced by performing a dot-product between a
filter and a same-sized portion of the input fmap's tensor. An example is shown
in the highlighted cells in Figure~\ref{fig:conv_background}, along with the
formula to compute each of the output fmap values. As one filter produces one
output feature map, the number of channels (feature maps) in the output is the
same as the number of filters (K). The number of output fmaps is the same as the
batch size (N).

\section{Convolution ABED Approach}
\label{sec:approach}

\begin{figure*}[tbp]
 \centering
	\includegraphics[width=2.1\columnwidth]{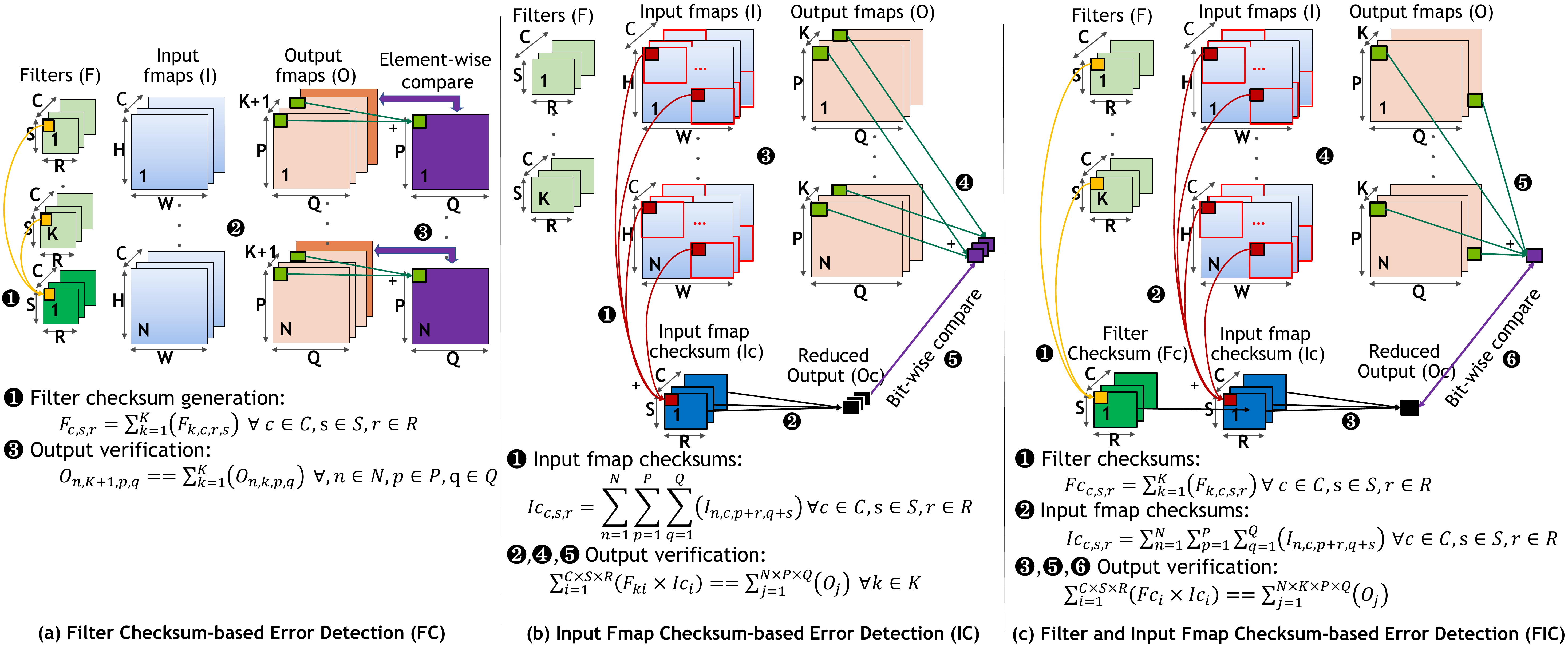}
	\vspace{-0.2in}
	\caption{A depiction of the filter-only (FC), input fmap-only (IC), and filter
	and input fmap (FIC) checksum-based error detection schemes.
	Formulae to generate the
	checksums and verify output fmap values are also shown. }
	\label{fig:abed}
\end{figure*}

Verifying every output value of a convolution might require duplicating the
entire operation. Instead, the focus of this work is on verifying just the reduced output, i.e.,
sum of all the output elements. This reduced output can be
computed from the inputs directly with far fewer computations. We essentially
use a different sequence of sums and products. Since integer sum and product
operations are commutative and associative, changing the order of the
operations is not a concern.  Based on this key insight, we explore the following three
schemes to verify a convolution, which are summarized in Figure~\ref{fig:abed}. 

\subsection {Filter Checksum-Based (FC)} 

In this scheme, 
a 3-D filter checksum tensor is computed by performing an element-wise
sum (using sum as a checksum function) across all the 3-D filter tensors 
(\mybtextcircled{1} in Figure~\ref{fig:abed}(a)).  
This new checksum filter is convolved
with the input fmaps to compute an extra output fmap, which is
used to verify the original fmaps' values. The original output fmaps'
values are reduced across the channel dimension to generate a reduced
fmap, which is compared element-wise for equality with the extra output
fmap for verification (\mybtextcircled{3} in Figure~\ref{fig:abed}(a)). 
This method protects the computation involved in the convolution, the data
storage and transportation of filters (between DRAM/L2 and registers) and
output fmaps, but not the storage and transportation of input fmaps.
This scheme increases the number of operations in the convolution by a factor of 1/K 
and introduces PQNK operations for output verification. Here PQNK refers to
P$\times$Q$\times$N$\times$K.  We omit the multiplication symbols while
referring to products of the parameters of the convolution for brevity.

\subsection{Input Checksum-Based (IC)} 

This scheme uses 
checksums of input fmaps, which can be computed in one of two ways: 
(1) summing input
fmaps' values element-wise across batches to add a new checksum batch, and (2)
summing elements of the portions of the input fmaps that are used to
perform the dot-product with the filters during the convolution operation to
produce a tensor that is the same size as the filter. 
		
The first option effectively increases
the batch size by one.  The original output fmaps'
values are reduced across all the original batches to generate a batch of
checksum output fmaps, which is compared for element-wise equality with
the extra batch of output fmaps generated using the checksum input
fmaps for verification.  This option is  attractive if the batch size is large
because the effective overhead of running the larger convolution would be
small.  However, for small batch sizes, which is common in safety-critical
systems~\cite{TeslaT4, HanMD15}, 
this option can result in high overheads. 

The second option reduces the input fmaps into a separate checksum tensor, which is 
the same size as a filter for the layer 
(\mybtextcircled{1} in Figure~\ref{fig:abed}(b)). 
This checksum tensor is then
convolved with K filters to produce exactly K values. 
The output fmaps are reduced across height, width, and batch dimensions and then compared
with these K values element-wise for equality for verification 
(\mybtextcircled{4} and \mybtextcircled{5} in Figure~\ref{fig:abed}(b)).
This method protects the computation involved in the convolution, the data
storage and transportation of input and
output fmaps, but not the storage and transportation of the filters. 
The number of additional computations needed for the convolution is CRSK,
input fmap checksum generation is PQNCRS, and output fmap checksum generation for
verification is PQNK.  

\subsection{Filter and Input Fmap Checksum-Based (FIC)}

The scheme, similar to the one proposed by prior work~\cite{Marty2018}, creates
checksums for both the filter and input fmaps (\mybtextcircled{1} and
\mybtextcircled{2} in Figure~\ref{fig:abed}(c)).  Using the two checksums, we
perform an extra convolution (\mybtextcircled{3} in Figure~\ref{fig:abed}(c)).
This operation can also be implemented as a vector-vector dot-product because
the filter checksum size is same
as the input fmap checksum size.  This operation produces a single value, which
is used to verify the original computation. The original convolution is run
with the original parameters and the output is reduced to a single value, which
is verified using the value generated by the dot-product.  (\mybtextcircled{4},
\mybtextcircled{5}, and \mybtextcircled{6} in Figure~\ref{fig:abed}(c)). 
%
This method protects the computation involved in the convolution and the data
storage as well as the transportation of filters and the
input and output fmaps.  The number of additional computations needed for
dot-product is CRS, input fmap checksum generation is PQNCRS, and output fmap
checksum generation for verification is PQNK.  

\subsection{Trade-offs} 
\label{sec:approach:tradeoffs}

Table~\ref{tab:impl_options} summarizes the trade-offs offered by the FC, IC,
and FIC techniques in terms of the number of tasks that must be performed
online and the protection they provide.  The table shows that FIC offers better
protection than FC and IC by protecting the storage and data movement of both
the filters and inputs.  The filter checksums can be generated offline because
the weights are known before a CNN is deployed.  However, the input and output
checksum generation and verification tasks must be performed online.  Since the
online tasks needed for the FIC and IC techniques are similar, the runtime
overheads also expected to be similar.  Given that FIC offers superior
protection compared to IC but the runtimes are expected to be similar, we do
not investigate IC further.

Since FC must run a larger convolution, the overhead can be higher than
FIC. However, FC can be faster if the larger convolution adds minimal
overheads, which is possible with the use of network pruning
techniques~\cite{Molchanov2016, Huang2018}. Network pruning
improves network performance by 
identifying and removing filters that contribute minimally to the accuracy of the
network.  With the use of pruned networks, the number of filters per layer may
diminish so that adding the checksum filter introduces minimal overheads
(explored further in Section~\ref{sec:results}). 

\begin{table}
\footnotesize
\caption{Trade-offs between the FC, IC, and the FIC techniques.
    Entries marked Yes/Offline and \textcolor{red}{No}/\textcolor{red}{Online} are favorable and
	unfavorable, respectively.}
	
\centering
\label{tab:impl_options}

\begin{tabular}{|m{0.5cm}|m{1.8cm}|m{1.4cm}|m{.75cm}|m{0.75cm}|m{0.75cm}|}
\hline

	& \multicolumn{2}{m{3.2cm}|}{Criteria} & FC & IC & FIC  \\
	\hline

	\multirow{5}{*}{\rotatebox[origin=c]{90}{\parbox[c]{1.2cm}{\centering Additional Work}}} & \multicolumn{2}{m{3.6cm}|}{Filter checksum generation} & Offline & - & Offline \\
	\cline{2-6}
	
	&\multicolumn{2}{m{3.6cm}|}{Input fmap checksum generation} & - & {\color{red} Online} & {\color{red} Online} \\
	\cline{2-6}

	& \multicolumn{2}{m{3.6cm}|}{Avoid running a larger convolution} & {\color{red} No} & Yes & Yes \\
	\hline 

	\multirow{3}{*}{\rotatebox[origin=c]{90}{Protects}} & \multicolumn{2}{m{3.6cm}|}{Computation} & Yes & Yes & Yes \\
	\cline{2-6}

	& \multirow{2}{2cm}{Storage and transportation} & Filters & Yes & {\color{red} No} & Yes \\
	\cline{3-6}

	& & Input fmaps & {\color{red} No} & Yes & Yes \\

	\hline 

\end{tabular}
\end{table}

\section{Implementation}
\label{impl}

This section addresses challenges that arise while implementing
ABED on a GPU-based system. Specifically, we explain (1) how to maintain high
coverage by avoiding overflow while using reduced-precision operations and
storage, the use of which is prevalent in inference platforms, (2) task-fusion
based optimizations/modifications we propose to the highly-optimized inference
platforms to minimize the overheads introduced by ABED, and (3) modifications
needed to the inference deployment frameworks for seamless integration with
ABED.

\subsection{Handling Reduced-Precision Operations}
\label{impl:prec}

The use of reduced-precision fixed-point data types has been explored
both in research as well as many commercial products. For example, 8-bit integer
arithmetic is supported in Google's Tensor Processing Unit, NVIDIA's
Volta and Turing GPUs, Intel Xeon Scalable Processors, and ARM
CPUs~\cite{GoogleTPU, VoltaArch, IntelNervana, ARMA64}. Fixed-point
arithmetic suffers from overflow if the result does not have sufficient bits to
represent the full value. Here we describe a method to ensure full error
coverage while using reduced-precision fixed-point data types.

Convolutions that use 8-bit integers (int8) store the filters and input fmaps
using int8 data types.  Each output fmap value is obtained by performing a
dot-product using one filter (of size CRS) with a same sized portion of the
input fmap, illustrated by the highlighted portion in
Figure~\ref{fig:conv_background}.  In this operation, CRS 16-bit values, each
of which is a product of two int8 values, are summed together.  Making
no assumption about the values, the result can be accurately represented using
$[16+log_2(CRS)]$-bit integers. For most practical values of C, R, and S
($CRS\leq64K$), int32 is sufficient to avoid overflow during convolutions. 

We use two's complement integer summation as the
checksum function. To avoid overflow during checksum generation, we use int32 operations. For
the FC technique, we store the int32 checksums as a tuple consisting of up to
four int8 values, creating up to four checksum filters.  No information is lost
with this scheme. The extra output fmap values are shifted left by $1$,
$8$, $16$, and $24$, respectively, and added together across the channel
dimension. These values are then compared with the output fmap checksums,
which are obtained by summing the original output fmap values across the channel
dimension (K additions). The reduced result can be accurately represented using
$[16+log_2(CRSK)]$-bit integers. For most practical values of C, R, S, and K,
64-bits are sufficient.

For the FIC technique, the filter checksum is obtained by summing K int8
filters and can be accurately represented by $[8+log_2(K)]$-bit integers.
Each input fmap checksum value is computed by summing PQN int8 input fmap
values, requiring up to $[8+log_2(PQN)]$ bits.   The result of the convolution
of the checksums would require up to $[16+log_2(PQNKCRS)]$ bits.  For most
practical purposes, int32 and int64 are sufficient to store the checksums
and convolution results, respectively. 

Table~\ref{fig:precision} summarizes the maximum number of bits needed to
accurately represent the values at different points during the convolution
operation for the FC and FIC techniques based on worst-case overflow
analysis. We assume unsigned integers for this analysis. The
requirements can be slightly less for signed integers. 
For example, the result of multiplying two signed 8-bit integers (with 1 sign
bit) can be accurately represented using 15-bit signed integers (with 1 sign
bit).
Since all parameters are known prior to a neural network deployment, the
ABED precision requirements can also be determined. For the networks used in this paper,
int64 checksums are sufficient to avoid overflow. If the precision requirements
increase beyond int64, modular arithmetic can be used to limit the highest
precision operation used by the verification kernel (to reduce runtime
overhead) with some loss in coverage, which we do not explore in this paper.

\begin{table}
\scriptsize
\caption{The bit requirements to accurately represent
	the results while verifying intb (e.g., int8 for b=8)
	convolutions. }
\centering
\begin{tabular}{|m{2.5cm}|m{2.6cm}|m{2cm}|}
	\hline
	& FIC & FC \\
	\hline
	Input fmaps	& b	& b	\\
	\hline
	Input fmap checksum & b+$log_2$(K) & -  \\
	\hline
	Filters		& b & b \\
	\hline
	Filter checksum & b+$log_2$(PQN) & b \\
	\hline
	Output fmap	& 2$\times$b+$log_2$(CRS) & 2$\times$b+$log_2$(CRS) \\
	\hline
	Reduced output fmap & 2$\times$b+$log_2$(PQNKCRS) & 2$\times$b+$log_2$(CRSK) \\
	\hline
	Dot-product output & 2$\times$b+$log_2$(PQNKCRS) & - \\
	\hline
\end{tabular}
\label{fig:precision}
\end{table}

Floating-point arithmetic suffers from both overflow and rounding. 
While we explored ways to address these issues, the discussion to
maintain high error coverage using floating-point data types is deferred to
Section~\ref{sec:discussion}, as commercial implementations increasingly
use fixed-point arithmetic due to its performance and energy advantages.

\subsection{Framework Modifications}

\begin{figure}[btp]
 \centering
	\includegraphics[width=\columnwidth]{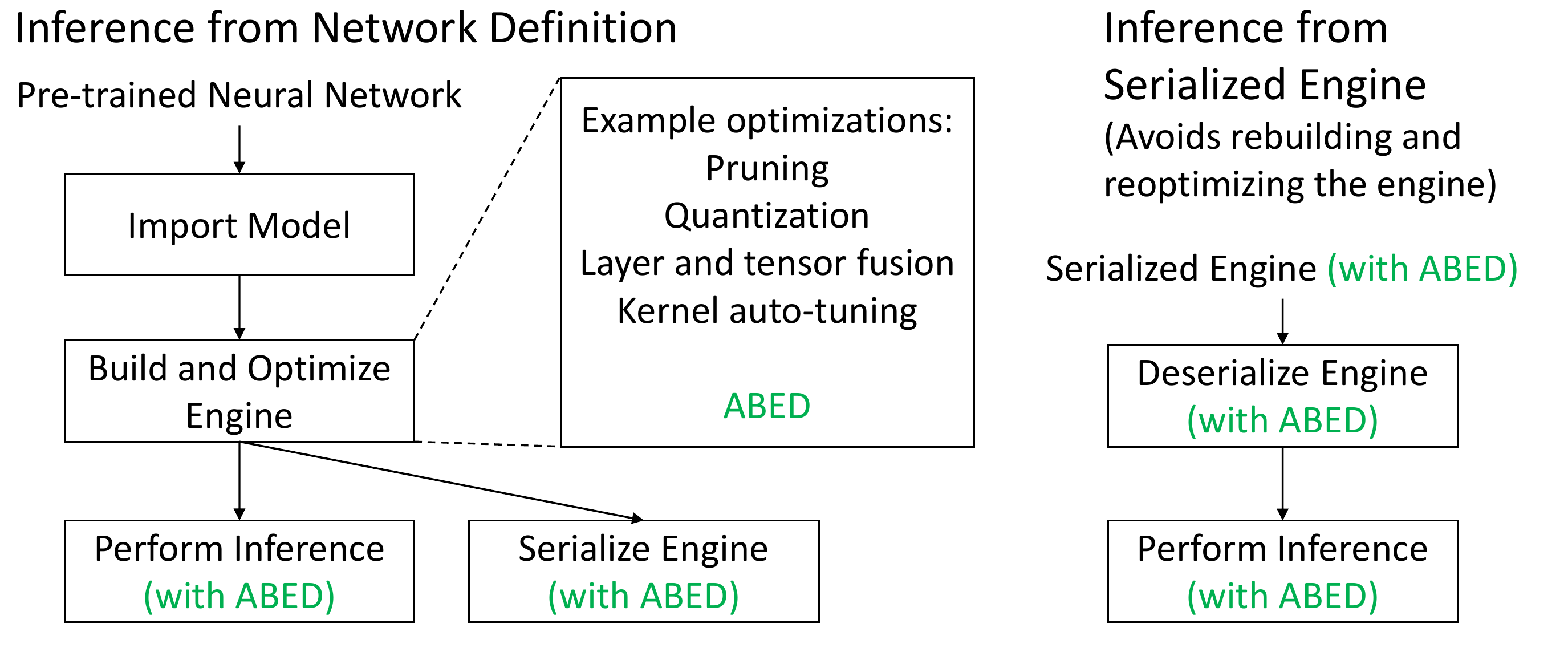}
	\vspace{-0.2in}
	\caption{Inference deployment steps and where ABED can be included for
	seamless integration.} 
	\label{fig:framework}
\end{figure}

Once a neural network is trained, it is optimized and prepared for deployment
using a platform for high-performance inference (e.g., TensorRT).  The
optimizations involve pruning, weight and activation precision calibration
(also known as quantization), layer and tensor fusion, and kernel auto-tuning.
These optimizations are typically performed once to create an inference engine,
which is then serialized to avoid preparation overheads. We perform the
following during the optimization step. (1) We create checksum filters and
store them along with the filter tensor and in separate storage for the FC and
FIC techniques, respectively. (2) We introduce all the additional online tasks
that should be executed during inference as part of the ABED scheme (e.g.,
input and output checksum generation). Figure~\ref{fig:framework} summarizes
the proposed changes for seamless integration of ABED in a tool-chain used to
deploy trained models for inference.
ABED is independent from all optimization described above, except the layer and
kernel fusion. We next describe how ABED can be applied to highly-optimized
convolutions that are fused with subsequent layers in CNNs. 

\subsection{Kernel Modifications}
\label{impl:mod_gpu}

\begin{figure}[btp]
 \centering
	\includegraphics[width=\columnwidth]{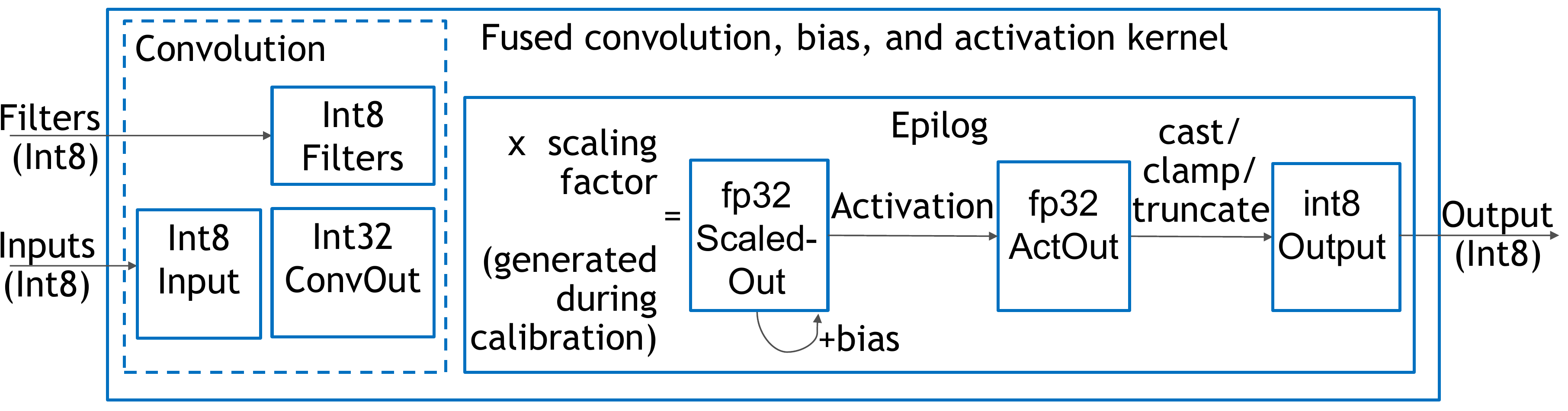}
	\vspace{-0.2in}
	\caption{The logical computation flow 
	in a fused
	convolution, bias, and activation kernel.} 
	\label{fig:int8conv}
\end{figure}

As described in Section~\ref{sec:background}, common convolution operations
take two 4-D tensors as inputs, one each for input fmaps and filters (I and F)
and produce a 4-D tensor of output fmaps (O).  Convolution, bias, and
activation operations are typically fused together for performance.  Such fused
operations perform $O=activation(conv(x)+bias)$.  For int8 convolutions, I and
F use int8, and O uses either int8 or fp32.  Figure~\ref{fig:int8conv} explains
the logical flow of computation within such fused kernels.  For int8
convolutions, the output of the convolution operation is an int32 result
(ConvOut in the figure). This intermediate result is then scaled using a
scaling factor that is generated during the calibration step, which produces an
fp32 result (ScaledOut in the figure).  This step assumes that the scaled int32
result can be accurately represented using an fp32 data type.  Bias is added to
ScaledOut. The activation function is then applied to it to produce ActOut
(another fp32 value).  If this value (ActOut) is too large to be accurately
represented using int8 data type, it will be clamped and truncated to
produce an int8 output value.  We refer to all the operations after the
convolution operation as {\em epilog}.

\begin{figure*}[ht]
 \centering
	\includegraphics[width=1.8\columnwidth]{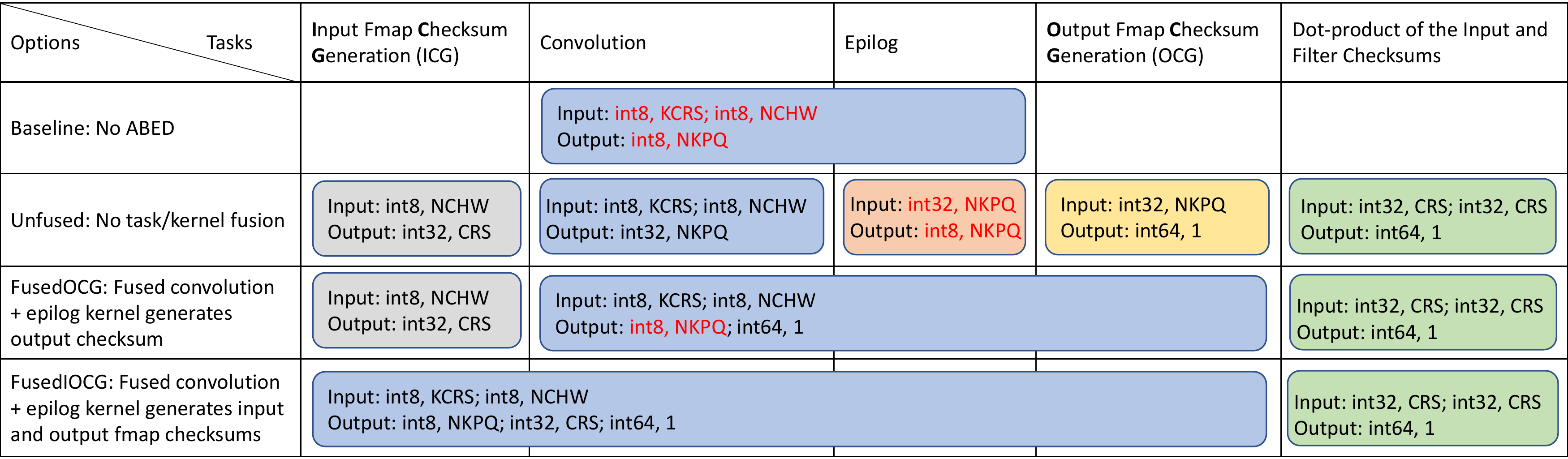}
	\vspace{-0.1in}
	\caption{Implementation options for the FIC technique are shown.
	Each colored box represents a GPU kernel and 
	shows the data type and size of the inputs and outputs.  Inputs/outputs
	for which the data transportation is not protected
	are shown in red. }
	\label{tab:impl_options_fic}
\end{figure*}

The ABED techniques verify just the result of the convolution. Hence, the
intermediate output (ConvOut in Figure~\ref{fig:int8conv}) must be verified
before the epilog is applied, which can be performed by either using
un-fused kernels or fusing some part of the output fmap checksum generation
task with the fused convolution + epilog kernel.
Figure~\ref{tab:impl_options_fic} lists some of the options for the FIC
technique.  For each option (one row in the table), we list the tasks that must 
be performed (in columns) and show the unfused/fused kernels that the option
will execute. For each of the kernels, we show the data types and sizes of its
inputs and outputs.

\noindent
{\bf FIC Technique:}
The following seven tasks (one more than what is listed in
Figure~\ref{fig:abed}(c)) must be performed for the FIC technique to detect
errors in the fused convolution + epilog kernel: (1) filter checksum
generation, (2) input fmap checksum generation (ICG), (3) dot-product of the
filter and input fmap checksums, (4) convolution operation, (5) output fmap
checksum generation (OCG), (6) epilog, and (7) verifying the output fmap
checksum with the output of step (3).  The first task is performed offline, as
described above. All other tasks are performed online.  Since task (7) involves
comparing just two values for bit-wise equality, it can be performed on the
host CPU. There are several ways to perform the remaining online tasks on a
GPU\@.  
A simple option is to use different GPU kernels to perform each of the tasks,
shown as option \UF{} in Figure~\ref{tab:impl_options_fic}.  The input checksum
generation kernel reads the input fmaps in int8 data type of size NHCW and
generates an int32 checksum vector of size CRS.  The convolution kernel reads
in the int8 filters and input fmaps of sizes KCRS and NCHW, respectively, and
writes out an int32 output of size NKPQ (ConvOut in 
Figure~\ref{fig:int8conv}). The next kernel reads ConvOut
and applies epilog to produce an int8 output of the same size as the input. The output
fmap checksum generation kernel reads the convolution output again to generate
a single int64 checksum value. Lastly, the dot-product kernel reads in filter
and input fmap checksums, and two int32 vectors of sizes CRS. It produces a single
int64 value, which is compared with the output fmap checksum for equality.
This implementation does not protect the epilog output and introduces several
additional data transfers including the convolution output stored in int32 data
type (4$\times$ the size of an equivalent int8 structure).

Task fusion can significantly reduce data movement. While we explored
multiple options to (partially/fully) fuse tasks, we describe only two
options here for brevity.  (1) The convolution, epilog, and output fmap
checksum generation kernels can be fused into a single kernel to limit
data movement introduced by ABED (\FR{} in
Figure~\ref{tab:impl_options_fic}). It generates a int64 value for output fmap
checksum along with the output of the fused convolution + epilog tasks, i.e.,
int8 output fmap tensor of size NKPQ.  (2) To further reduce data movement and
provide coverage for epilog's output, the output checksum generation task can
be modified to produce the input fmap checksum for the subsequent layer, if the
next layer is a convolution layer (\AF{} in
Figure~\ref{tab:impl_options_fic}). It essentially
fuses input checksum generation task with the fused convolution + epilog +
output checksum generation kernel. This optimization duplicates the epilog, but
we assume that data movement saving will improve runtime more than the
overhead introduced by duplicating compute-bound epilog. 
The \FR{} and \AF{} options require changes to the existing fused convolution +
epilog kernel offered by frameworks such as cuDNN and TensorRT.

\noindent
{\bf FC Technique:}
The following are the tasks needed for the FC technique (one more than what is
listed in Figure~\ref{fig:abed}(a)): (1) filter checksum generation, (2)
convolution operation, (3) epilog, and (4) output fmap reduction and
verification.  The first task is done offline. 
The verification task involves computing checksums from the original and extra
output feature maps across the channel dimension separately and comparing them
for bit-wise equality. This task must be performed before the epilog and can be
implemented in multiple ways. 
The first option is to not fuse the operations and launch
separate kernels for the convolution, epilog, and output fmap checksum
generation and verification. This option is similar to \UF{} for FIC. 
Fusing the output fmap checksum
generation and verification task with the already fused convolution + epilog
kernel will reduce data movement. This option is similar to \FR{} for FIC. 

Since this technique adds checksum filters, the runtime of the
larger convolution can be higher. The increase can be super-linear
in the number of filters for certain architectures and convolution parameters.
The convolution implementation is often tiled. The runtime may not increase if
a tile boundary is not crossed and can increase significantly if it is crossed.
Coordination with the pruning techniques may help in reducing this overhead.
When the network is being prepared for
deployment, filter-pruning is commonly employed to optimize the network's
performance and storage~\cite{Molchanov2016, Huang2018}.  
Reducing the number of filters at this step
such that the checksum filters can be included without introducing a new tile
of work can reduce the overhead significantly. 

The output of the fused operation must be trimmed such that the extra
feature maps are ignored by the subsequent layer.  Trimming can be fused with
the output verification task.  We do not explicitly study the effects of
trimming because implementing it simply requires skipping some writes.

\section{Evaluation Methodology} 
\label{sec:method}

We evaluate the overheads introduced by the ABED techniques to the
convolutional layers from different CNNs. We use VGG16, ResNet18, and ResNet50
for analysis~\cite{SimonyanZ14a, HeZRS15}. We evaluate using two different
image sizes---224x224, the size of the images in the ImageNet
dataset~\cite{imagenet_cvpr09}, and 1080x1920, the resolution of the images in
a full-HD or 1080p video.  

\subsection{Compute/Data Movement Overheads Estimation}
\label{method:model}

We first analytically evaluate the increase in the compute and data
movement operations when we apply the FC and FIC ABED techniques to the networks.
In this analysis, we abstract away implementation details and only consider the
arithmetic operations such as multiplication, addition, fused multiply-addition (FMA),
activation, and type-casting. Similarly, we also count the bytes of data that
form the inputs and outputs of different implementations for the FC and FIC
techniques, as listed in Section~\ref{impl:mod_gpu}. 

\subsection{Runtime Overhead Evaluation}
\label{method:runtime}

We experimentally evaluate the runtime of convolutions by creating a cuDNN-based workload that
sets up, initializes, and runs convolutions in a loop. We compile this workload
using CUDA 10  and use cuDNN
7.3~\cite{cuDNN:Online} on both a Jetson AGX Xavier system and an x86-based desktop
with a V100-based GPU (Titan V)~\cite{VoltaArch, JetsonXavier:Online}. 
For performance analysis, we lock the CPU, GPU, and memory clocks on the Jetson
board and lock the application clocks on the V100 GPU.  We run the convolution
(and other operations needed by the ABED techniques) 200 times, recording the
average.  Since real-time applications (including safety-critical systems) use
small batch sizes, we use batch size of two on Jetson and eight on V100-based
system~\cite{TeslaT4, HanMD15}.  We ignore the first layer in each network
because it is not well optimized by cuDNN.  We use NHWC memory layout to tensor
storage, as int8 cuDNN convolutions are optimized for this layout.

For a baseline, we invoke the fused convolution and epilog kernel provided by
cuDNN (called cudnnConvolutionBiasActivationForward).  We implement versions
that are similar to \UF{} for the FIC and FC techniques.  As cuDNN does not
offer a convolution kernel that takes in two int8 tensors and produces an int32
tensor as output, we employ a version that produces fp32 output. The epilog,
when performed separately, invokes two GPU kernels (one each for adding bias
and applying activation) and generates a fp32 tensor as output (instead of an
int8 tensor, as mentioned in Section~\ref{impl:mod_gpu}, due to cuDNN API
limitations).  For analyzing the \UF{} implementation options for ABED, we also
collect results with just the unfused convoltuion and epilog kernels (without
ABED). This version launches one kernel each to perform the convolution
operation,
add bias, 
and apply activation.
We refer to it as the unfused baseline.

The checksum generation kernels are written in CUDA. Since the checksum
generation has similarities to the reduction operation, we use
previously-established optimizations. Specifically, we use the functions and
primitives (such as DeviceReduce and WarpReduce) provided by the {\em CUB} library
optimized implementations~\cite{cub}.  We try to minimize the
use of atomics 
and leverage faster memory (e.g., registers and shared memory) as much as
possible. We ensure that global loads are coalesced across warps and use wide
loads per thread (e.g., LD.128). We avoid control flow and use the dot-product
instruction (DP4A) whenever possible to avoid a compute
bottleneck~\cite{NVIDIA:ISA}. We specialize kernels for filter sizes 1 and 3,
strides 1 and 2, and for data types int8 and int32.
For the FC technique, we add 8 filters (4 for checksums and 4 with zeros for
int8) because the cuDNN version we use on our target device chooses efficient
kernels when the number of filters is a multiple of 8. We obtain the aggregate
runtimes per network by adding all of the GPU kernel runtimes and show the
runtime relative to one of the baselines mentioned above.  

\noindent
{\bf Effect of Task Fusion-based Optimization:} Since we could not modify the
closed-source cuDNN kernels to fuse checksum generation and output
verification tasks with the convolution for the \FR{} versions, we model the
runtimes.  We write separate CUDA
kernels to capture the overheads associated with performing the additional work
that will be fused with the convolution + epilog kernel. For FIC-\FR{}, the new
kernel fully reduces the output and writes out just one int64 value.  Since
using atomic operations for reduction can be a performance bottleneck, we
hierarchically reduce the output similar to the prior optimized reduction task
implementations~\cite{cub}.  For FC-\FR{}, the new kernel reduces the values
across the channel dimension for verification and sets a flag on an error
detection. We measure the runtimes of these kernels by running them on
silicon and add the runtimes to Baseline-Fused as an estimate of Kernel 3's
runtime (from Figure~\ref{tab:impl_options_fic}).

\subsection{Overhead Analysis of a Traditional ABFT Technique}
\label{method:abft}

As convolutions can be implemented using GEMM, ABFT can be employed at the
GEMM-level to provide protection. While prior work explained the resilience
benefits, no performance assessment was included~\cite{Santos2019}. We quantify
the overheads associated with the ABFT technique for GEMM and highlight the
benefits of our ABED solutions to protect convolutions.  An ABFT
approach performs the following tasks: (1) allocate larger input and
output matrices with space for checksums, (2) copy data from the original
matrices to the new locations, (3) generate checksums for both the input
matrices, (4) run the larger GEMM, instead of the original GEMM, (5) generate
both row and column checksums for the output 
(by reading the output matrix twice) 
and compare them with the extra row and column values generated by the GEMM,
and (6) copy the output matrix to the original location. We measure overheads for 
tasks (2)-(6) by developing CUDA kernels. We compare the
runtime of our implementation of the checksum generation task with the reduction
operation provided by CUB~\cite{cub} and use that whenever it is faster (assuming
checksum generation can be as fast as reduction). 
ABFT can be implemented to store the input checksums separately 
to avoid copying large matrices. Such an
implementation launches separate kernels to perform original GEMM and
generate extra output row and column via vector-matrix products, which
can be as slow as the original GEMM for some matrix sizes (used in CNNs). Due 
to such overheads, we do not explore this option in this paper. 

\subsection{Resilience Evaluation} 
\label{method:resilience}

We evaluate the resilience improvements offered by the ABED techniques using
three methods---analytical modeling, input-output error injections, and
accelerated-particle beam experiments.  The first method uses the same model
used above to analyze the increase in compute and data movement operations.
Here we analyze the fraction of compute and data movement operations that are
protected by the ABED techniques.
For the second method, we run the second convolution layer from ResNet18 with
input (fmap and filter) values initialized to 1.  We perform three
error injection campaigns to study the effect of injecting single bit-flips
into input fmaps, filters, and output fmaps. For each experiment
in a campaign, we randomly flip a bit in a randomly chosen location in the
tensor and study whether the ABED approach detects the error. 

We conducted two accelerated-particle beam experiments at ChipIr at Rutherford
Appleton Laboratory and ICE-II at LANSCE~\cite{chipir, nowicki2017alamos}
using our implementations of the \UF{} option 
for the FIC and FC techniques and baseline. We excluded epilog in these experiments. 
The input tensors were initialized to 1.
After each convolution we verified the output with the expected golden values
that were collected during fault-free runs.  Any mismatch was recorded as an
SDC. 
Our investigation suggests that the output tensor was corrupted for some runs
after the ABED checks were completed, likely when the output was being reduced
to compare against the golden checksum to determine the SDC. These output
corruptions are out of the coverage scope for ABED and hence we do not consider
them as SDCs in this study.
We also recorded whether our ABED scheme was able to detect the error. We
tested with NVIDIA Quadro GV100 GPUs with HBM2 ECC always On, and on-chip ECC
On and Off.

\section{Results}
\label{sec:results}

\subsection{Compute/Data Movement Overheads Estimation}

\begin{figure}[tbp]
 \centering
	\includegraphics[width=\columnwidth]{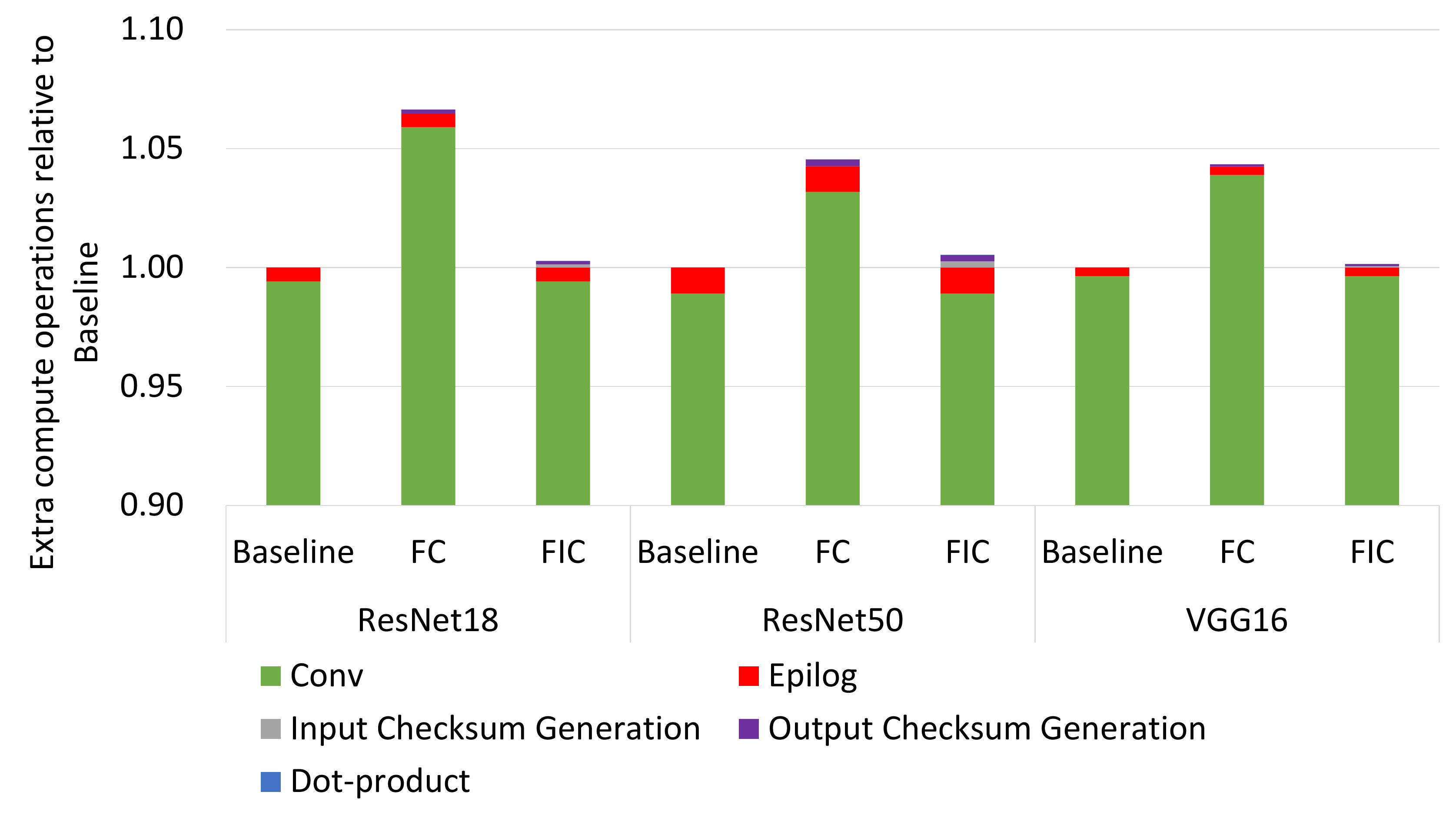}
	\vspace{-0.25in}
	\caption{The increase in the number of logical compute operations for
	the two ABED techniques relative to the baselines. }
	\label{fig:sol-compute}
\end{figure}

We first study the increase in the logical compute and data movement operations
based on the model described in Section~\ref{method:model}.
Figure~\ref{fig:sol-compute} shows a breakdown of the number of arithmetic
operations in convolution, epilog, checksum generation, and 
dot-product of the checksums for the baseline and the FC and FIC techniques.
The average increase in the number of operations is
small,  $<$7\% for FC and $<$1\% for FIC for the studied networks compared to
the respective baselines. 
The increase is relatively higher for the FC technique because it increases the
size of the convolution, unlike the FIC technique.  Results show that the extra
computations added for checksum generation and performing the dot-product of
the checksums are significantly less than 1\%. 

\begin{figure}[tbp]
 \centering
	\includegraphics[width=\columnwidth]{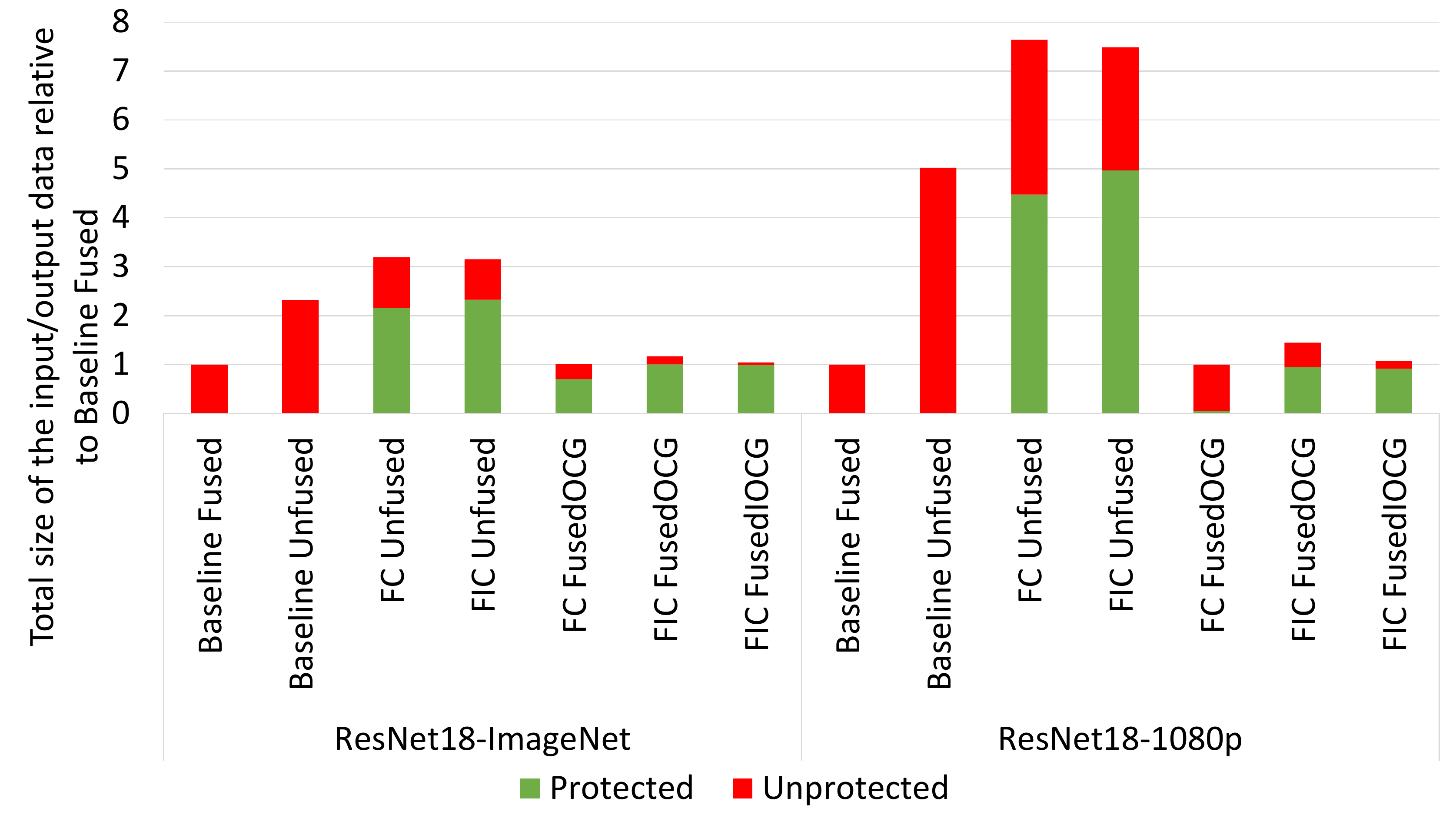}
	\vspace{-0.2in}
	\caption{The relative increase in the data that forms input and output
	of all the kernels for different implementations of the ABED techniques
	is shown here.
	} 
	\label{fig:sol-data}
\end{figure}

Different implementations listed in Section~\ref{impl:mod_gpu} perform the same
logical compute tasks, but differ in terms of data movement.
Figure~\ref{fig:sol-data} shows the bytes of data that form the inputs and
outputs of the different implementation options 
for ResNet18 using two input image sizes.  Results for other networks and input
sizes show similar trends (not shown for brevity).  The figure shows that the
fused versions transport significantly less data compared to the versions that
do not fuse tasks. 
Introducing separate kernels for checksum generation and verification
introduces more data movement, as is the case for FIC \UF{} and FC \UF{}.
Results also show that the FC \FR{} requires less data movement than FIC \FR{},
but FIC \FR{} offers better data movement protection by using input and weight
checksums, while FC \FR{} protects just the weight storage and movement.

\subsection{Runtime Overheads}
\begin{figure}[tbp]
 \centering
	\includegraphics[width=\columnwidth]{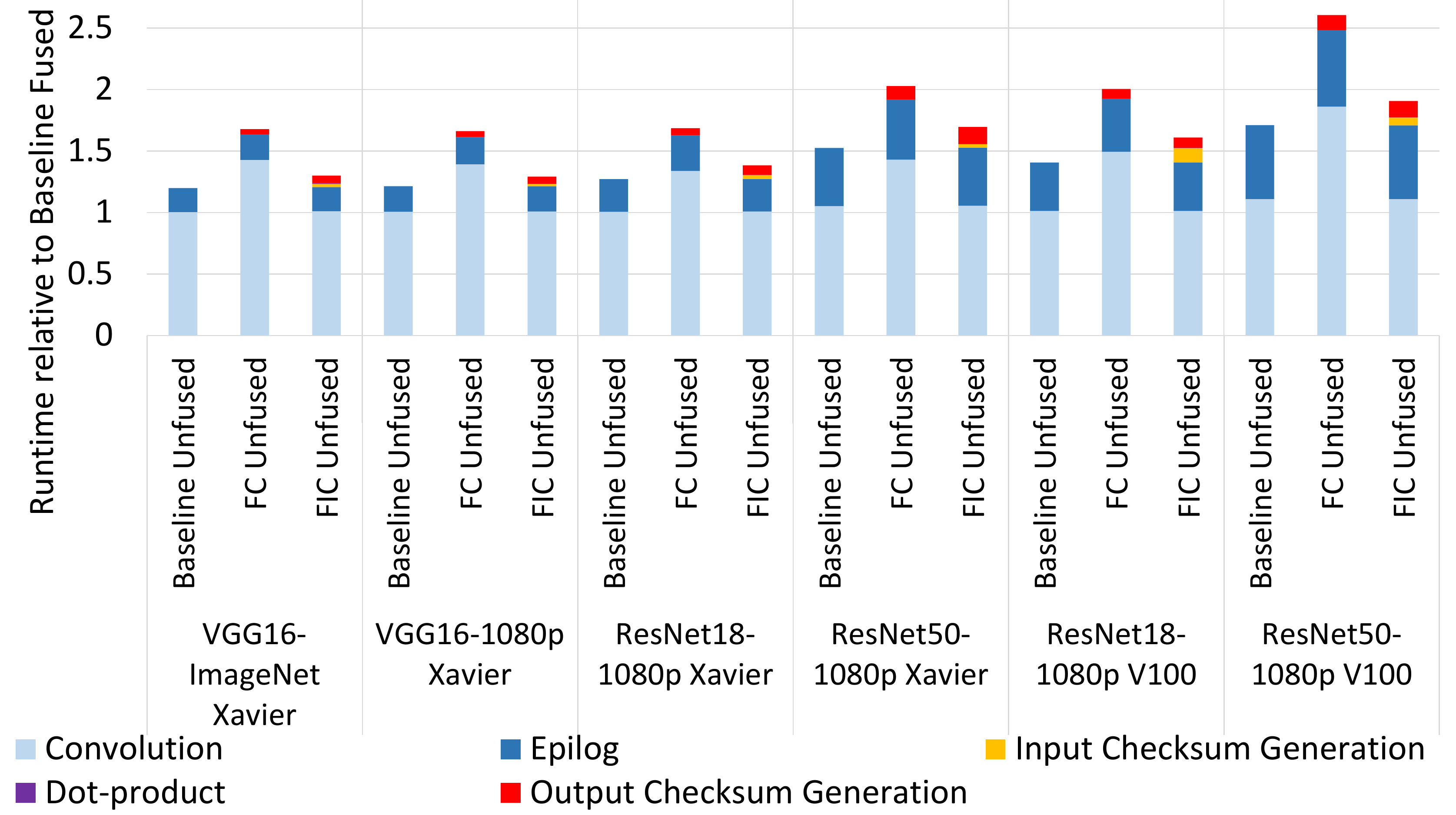}
	\vspace{-0.2in}
	\caption{The runtimes of the Unfused versions of the baseline and FC
	and FIC techniques for different neural networks are shown here.  } 
	\label{fig:int8-conv-1}
\end{figure}

Figure~\ref{fig:int8-conv-1} shows the average runtimes of the Unfused
options of the baseline and FC and FIC techniques.  
Results for the three networks using 1080p inputs are shown here. The results
are normalized to the Fused baseline. 
The runtime overheads for the FC technique stem
from running a larger convolution with additional checksum filters and output
checksum generation. For the FIC technique, the overheads stem from running the
input and output fmap checksum generation tasks and the dot-product of the
filter and input fmap checksums. 

\noindent 
{\bf FC vs.\ FIC:}
Results show that the runtime overhead introduced by the output checksum
generation is similar between the FC and FIC techniques.  The dot-product
kernel used during the FIC technique introduces negligible overhead across 
all the studied networks and architectures. The difference in overheads
between the FC and FIC techniques is mainly due to running the larger convolution versus
generating input fmap checksums online. The former introduces
higher overheads for all the networks and architectures we studied
(Figure~\ref{fig:int8-conv-1}). 
The results show that the overheads introduced by the checksum generation and verification tasks are small (4-20\%). The overhead introduced by the separate data-movement-heavy epilog is high which can be avoided by the task-fusion-based implementations discussed below. 

\noindent 
{\bf Model-specific Sensitivities:}
The overhead of output checksum generation for ResNet50 is 
higher compared to VGG16 and ResNet18. A primary reason for this difference is
that the overhead for verifying 1x1 convolutions is much higher compared to
verifying 3x3 convolutions, and ResNet50 uses many 1x1 convolutions while
ResNet18 and VGG16 do not use any.  The fraction of the work (and data
movement) performed for checksum generation to that of the baseline convolution
is much higher for 1x1 convolutions compared to 3x3 convolution. 
Fusing the output checksum generation task with the convolution operation can
help reduce the overheads. 

\noindent 
{\bf Architecture-specific Sensitivities:}
We show the results obtained from Jetson AGX Xavier (with batch size of two)
and a V100-based GPU (with batch size of eight) in Figure~\ref{fig:int8-conv-1}.  
The V100-based GPU offers approximately 10$\times$ 
compute throughput and 5$\times$ memory bandwidth 
compared to the Xavier GPU. As expected the baseline work is 
signficantly faster on the V100-based GPU. Since the throughput to bandwidth ratio
is higher for the V100-based GPU, overheads of the memory-bound tasks such as
checksum generation and epilog are higher. In fact, the tasks that are not
memory-bound in Xavier become memory bandwidth/latency limited in V100. 

The runtime overhead of generating input checksums is significantly lower than
generating output checksums. The main contributing factor is that the input
fmaps are 4$\times$ smaller compared to the non-scaled 32-bit integer output
fmap values. One exception to this finding is ResNet-18 with image size of
1080x1280 on the V100-based GPU. Input checksum generation for ResNet-18 incurs
high overhead because it becomes memory-bandwidth limited on this GPU. 

\noindent 
{\bf Input-specific Sensitivities:} 
Figure~\ref{fig:int8-conv-1} shows the relative runtime for VGG16 with
different input sizes (224x224 vs.\ 1088x1920). Since no significant difference
is observed, we do not analyze results with smaller image size for other networks. 

\begin{figure}[btp]
 \centering
	\includegraphics[width=\columnwidth]{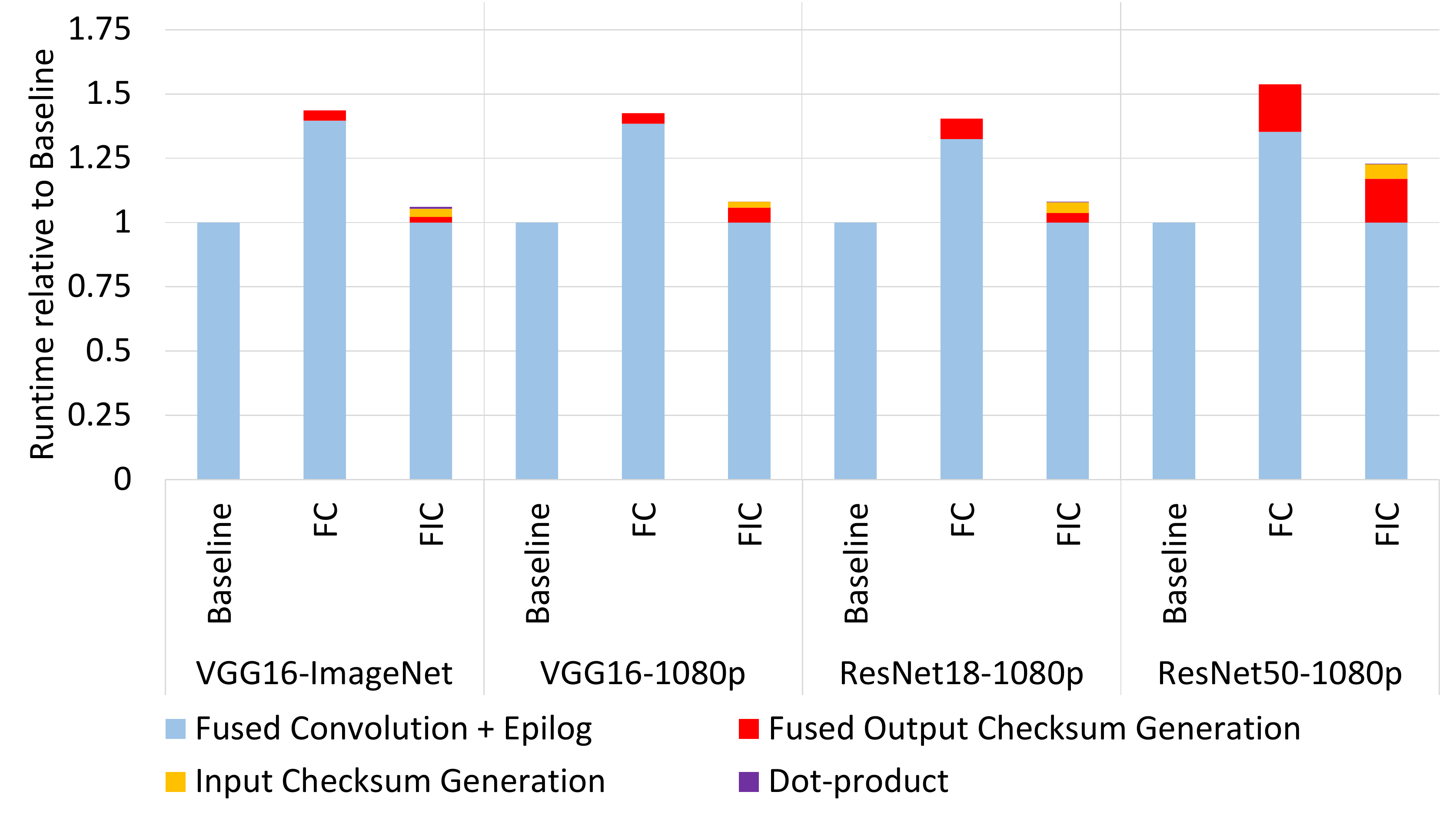}
	\vspace{-0.2in}
	\caption{The runtime overheads of \FR{} options for the FC and FIC techniques relative to the fused convolution + e05ilog kernel (Baseline) are shown here.} 
	\label{fig:fic-model}
\end{figure}

\noindent 
{\bf Effect of Task Fusion-based Optimization:}
The runtime overhead results for the \FR{} optimization for FC and FIC techniques 
obtained using the methodology described in Section~\ref{method:runtime} are
shown in Figure~\ref{fig:fic-model}. These experiments were run on Jetson
Xavier. These results suggest that task fusion is highly effective in reducing
the overheads by reducing memory traffic associated with epilog and additional ABED
tasks. It shows that the inference-level overheads for the FIC technique (6-23\%) 
are far lower than full duplication. The overheads for
the FC technique are higher mainly due to running the larger convolution (with
additional checksum filters). 

\begin{figure}[tbp]
 \centering
	\includegraphics[width=\columnwidth]{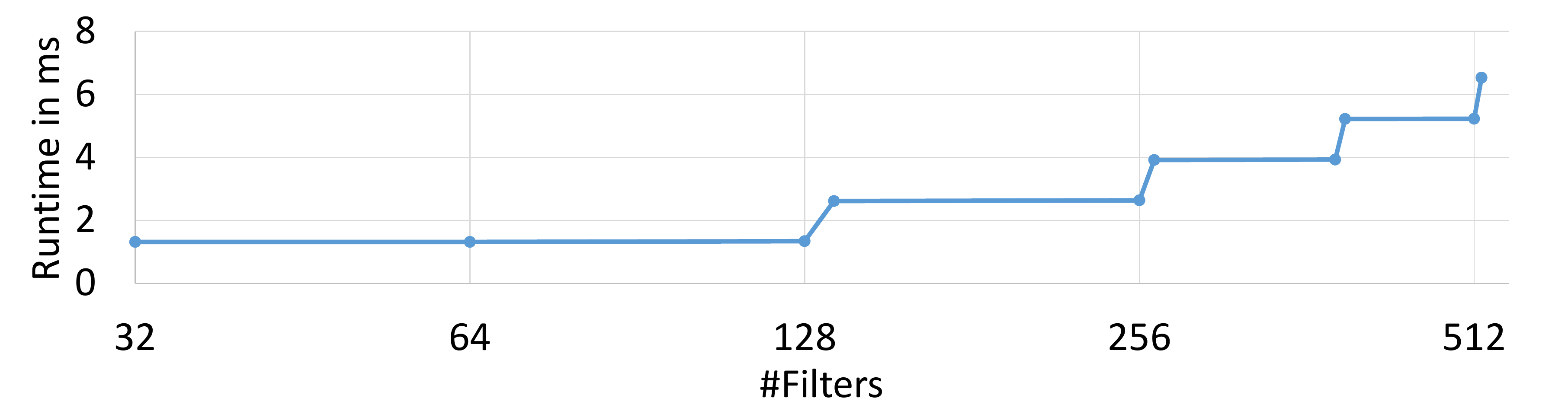}
	\vspace{-0.25in}
	\caption{Convolution runtimes with varying filter counts.} 
	\vspace{-0.1in}
	\label{fig:conv-trends}
\end{figure}

\noindent
{\bf Reducing overheads for the FC technique:}
In our FC implementations, we increase the filter counts by 8 (as described in
Section~\ref{method:runtime}). The runtime of the convolution, however,
increases disproportionately for some layers. We illustrate this behavior by
executing a convolutional layer with a varying number of filters.
Specifically, we ran 
an int8 convolution with
112x112 input fmap, 3x3 filters, 64 input channels, and stride and padding of 1. We 
vary the filter counts and show that adding just eight filters can introduce
up to 2$\times$ overhead. Figure~\ref{fig:conv-trends} shows these results.
Since cuDNN uses GEMM as a method to perform the int8 convolutions and GEMMs
use tiling, the sharp increase in the runtime is likely
due to the use of an additional tile. While such tiling effects are not a concern
for GEMMs with large dimensions, they can be problematic for the commonly used
convolutional layer dimensions (where K is small).

\begin{figure}[tbp]
 \centering
	\includegraphics[width=\columnwidth]{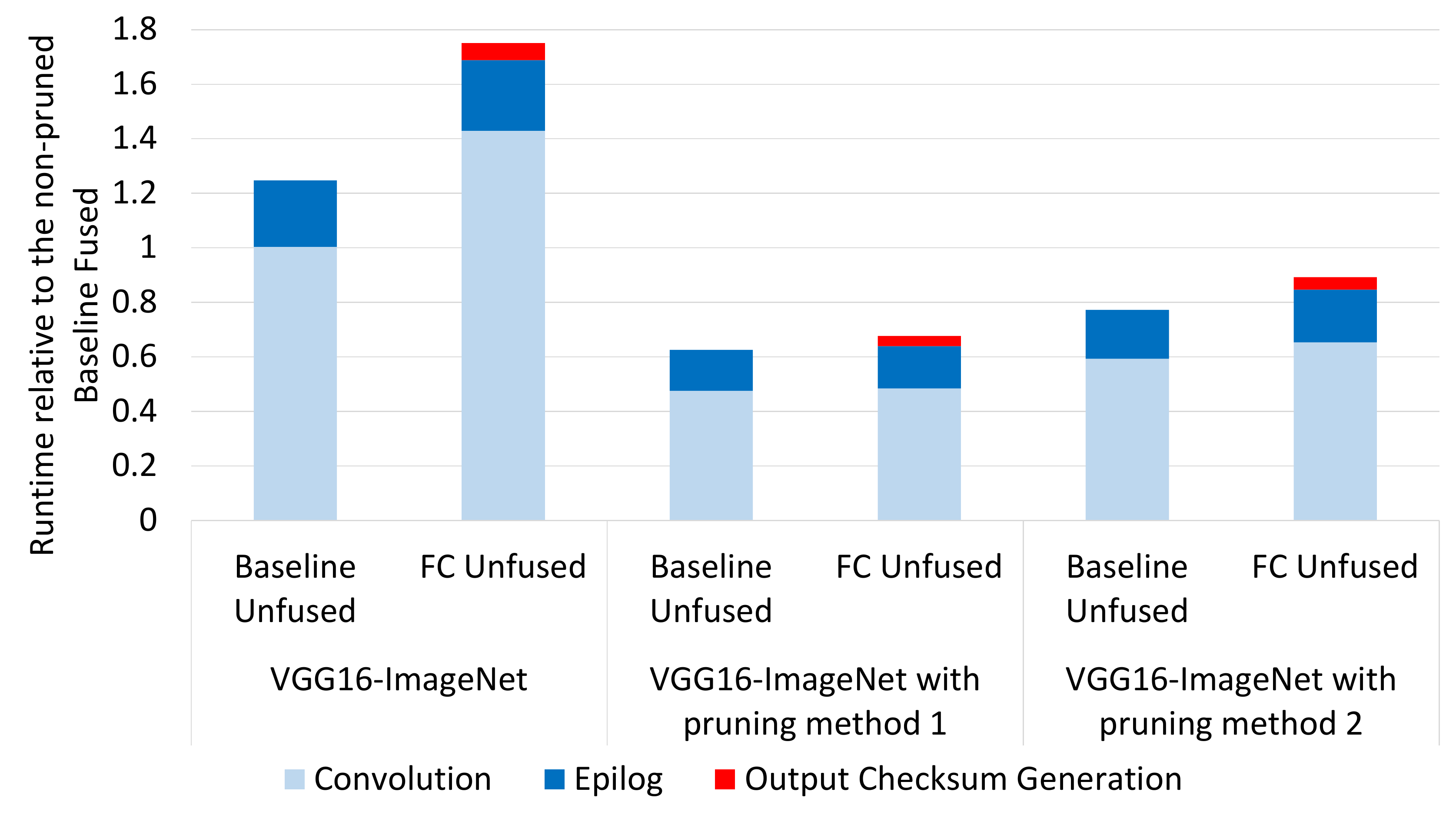}
	\vspace{-0.2in}
	\caption{The runtimes for VGG16 and two pruned versions
	for the Unfused baseline and FC technique are shown here. 
	}	
	\label{fig:fc-pruned}
\end{figure}

Instead of increasing the filter counts of the baseline network, creating space
for the filter checksums can eliminate this overhead.  As mentioned in
Section~\ref{sec:approach:tradeoffs}, network pruning techniques, which are
being adapted as a way to improve network performance, may create space for
filter checksums. These techniques identify and remove filters that contribute
minimally to the accuracy of the network. With the use of pruned networks, the
number of filters per layer may reduce even after adding the checksum filter,
the sharp increase in the convolution runtime can be avoided.  To test this
hypothesis, we conducted an experiment for the FC technique using VGG16 and two
pruned versions of the network. 
We obtain the number of pruned filters per
layer from a previously published result. Huang et al.,
studied two methods to prune the network~\cite{Huang2018}. The first approach ranks filters on
per layer basis, while the second ranks them across all the network.  
Our results in
Figure~\ref{fig:fc-pruned} demonstrate that the overheads from running the
larger convolution become small, 2\% or 10\% for the two pruned versions,
respectively, compared to the 42\% overhead for the non-pruned version. 

\begin{figure}[tbp]
 \centering
	\includegraphics[width=\columnwidth]{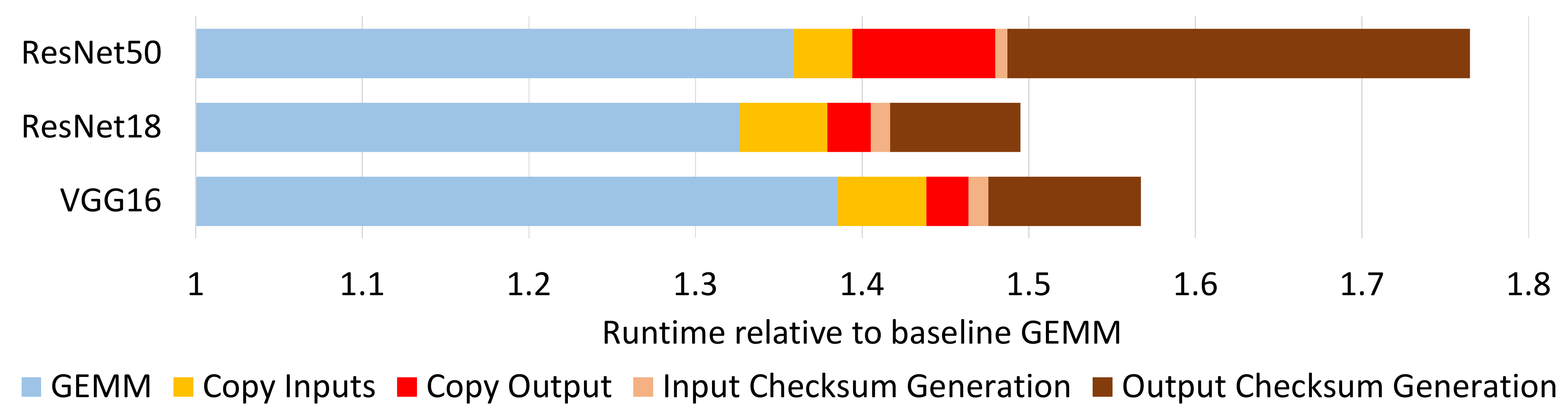}
	\vspace{-0.2in}
	\caption{Runtime breakdown for an ABFT implementation.} 
	\label{fig:abft_gemm}
	\vspace{-0.1in}
\end{figure}

\subsection{Overhead Analysis of a Traditional ABFT Technique}
\label{results:abft}

As convolutions can be implemented using GEMM, ABFT for GEMM (a well known
technique) can provide high protection to CNNs.  A recent study explained the
resilience benefits of such an approach~\cite{Santos2019}.  
In this section, we quantify the
overheads associated with the ABFT technique for GEMM, as described in
Section~\ref{method:abft}, and highlight the benefits of our ABED solutions to
protect convolutions in CNNs. Results
for the three networks for 1080p input are shown in Figure~\ref{fig:abft_gemm}.
These results show that running a larger GEMM
incurs high overhead (similar to the FC technique). This overhead
can be reduced using pruned networks (not explored here). As our ABFT
implementation embeds the row and column checksums along with the input
matrices to perform a larger GEMM, online data management (i.e., copying input
to larger matrices) introduces significant overheads. This overhead can be
avoided by allocating larger matrices in the first place for the inputs and
output with broader application knowledge and framework support, which is what
we propose for the FC technique. The FIC technique avoids running the larger
convolution altogether, simplifying data management.  Lastly, processing output
matrix twice to generate both the row and column checksums for error correction
capability can also introduce high overheads. Optimized implementations
that process the output just once will reduce this overhead. By focusing on
error detection alone, ABED significantly speeds up this step by generating a
single checksum.

\subsection{Resilience Evaluation}
\label{res:resilience}

{\bf Scope of protection:} Figures~\ref{fig:sol-compute} and~\ref{fig:sol-data}
show the scope of protection offered by different the FC and FIC techniques. The ABED 
techniques protect computation in the
convolution, input and output checksum generation, and dot-product kernel.  
Other than
convolutions, CNNs include activation, pooling, and fully-connected layers.
Activation layers are typically merged with the convolution layer and
they constitute a small fraction of the total compute (0.6\% for ResNet18
 and 1.1\% for ResNet50 as shown in
Figure~\ref{fig:sol-compute}). Only a few pooling layers are typically used in
CNNs (just two in ResNet18 and ResNet50, for example) and their compute
requirement is also small ($<$0.3\%). Fully-connected layers can be
converted to GEMMs and checksum-based ABED techniques can be applied for their
protection. For full network protection, ABED can be applied to convolutions
and fully-connected layers, and duplication can be used to protect the rest,
which constitutes a small fraction of the total compute, $<$1.4\% for ResNet18
and ResNet50. Compared to full inference-level duplication, the ABED approach
offers overheads that are lower by $>$4$\times$. 

The amount of protection offered by ABED for data storage and movement is
important for architectures that do not protect (with ECC/parity) storage and
transportation structures sufficiently against transient, intermittent, and
permanent errors. We show the levels of protection ABED techniques offer for
data movement in Figure~\ref{fig:sol-data}. Since FIC technique protects the
input data to the original convolution kernel, it provides better data storage
and movement protection than the FC technique. As the input fmap size
increases, the coverage offered by the FC technique reduces (compare FC \FR{}
results between ResNet18-ImageNet and ResNet18-1080p). While the coverage also
reduces for the FIC technique, the reduction is much less.  Results show that
FIC \AF{} offers highest data storage and movement coverage among all the FIC
options.

ECC/parity deployed in architectures used in HPC and safety-critical
systems provide no protection to computational units, one of the major sources
of intermittent and permanent errors. ABED techniques provide very high
protection for computational units along with storage and data transportation
protection.

\noindent 
{\bf Error injections:} We perform error injections into the input and output
tensors as described in Section~\ref{method:resilience}. Our results for the FC
technique show that all single-bit injections into non-zero filters and output
fmaps are detected by the ABED technique and no single-bit injections into
input fmaps are detected, as expected. A similar experiment for the FIC
technique shows that errors in the input fmaps, filters, and output fmaps are
detected.

\begin{figure}[tbp]
 \centering
	\includegraphics[width=0.95\columnwidth]{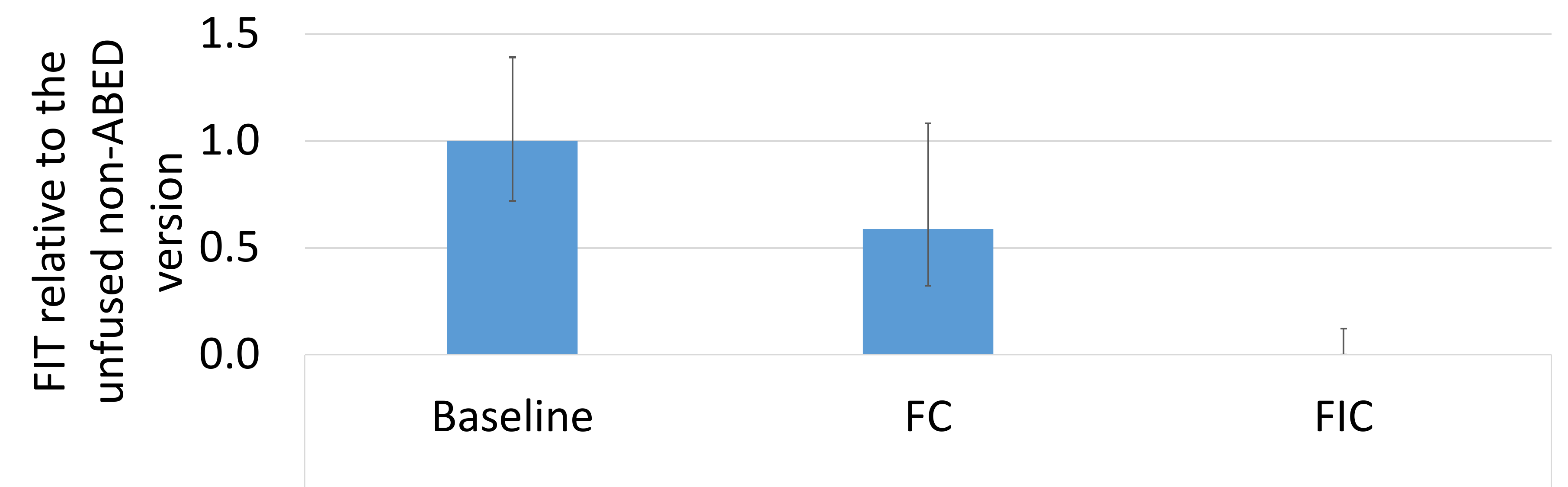}
	\vspace{-0.15in}
	\caption{The SDC FIT rate improvement with the FC and FIC techniques 
	without on-chip ECC.} 
	\label{fig:sdc-fit}
	\vspace{-0.1in}
\end{figure}

\noindent 
{\bf Beam testing:} To accurately quantify the vulnerability improvement, we
conducted accelerated-particle beam experiments as outlined in
Section~\ref{method:resilience}.  For the on-chip ECC Off experiments, the
results show a clear SDC FIT rate reduction trend for the FC and FIC
techniques.  We observed some SDCs when the FC technique was employed. Up on inspecting
the SDCs that were not detected by the FC technique, we found the output to
be corrupted such that the values in original fmaps when reduced to be compared
to the checksums result in no error detection (both the original output and
checksum values were corrupted).  Such manifestations indicate corruption to
the input fmap, which is not covered by this technique.  With on-chip SRAM
(register file, L1/L2 cache) ECC/parity protection enabled, the likelihood of
such errors will be low.  We observed no SDCs while running convolutions with
the FIC technique.  The ABED techniques detected a few errors when no SDC was
observed. These extra errors could be the result of a fault in the verification
kernel.  Figure~\ref{fig:sdc-fit} shows the FIT rate improvement results with
on-chip ECC Off.

With on-chip ECC On, we expect both the FC and FIC techniques to offer high and
comparable protection. Due to the cost of experimentation and limited
availability of the beam time, we verify this only for FIC.  In this
experiment, the FIC technique detected all observed SDCs.  The error bars for
this experiment were relatively large, however, due to limited beam time and
relatively lower SDC rate of the GPU (compared to on-chip ECC Off).

\section{Discussion: Managing Rounding Error}
\label{sec:discussion}

While commercial systems increasingly use fixed-point data types during
inference, the use of half-precision floating-point data type (fp16) is also
common. All the techniques described in the paper are applicable to fp16
convolutions.  Due to the non-associative nature of floating-point
operations, the final comparison 
cannot be exact. 
We can test whether the two reduced values computed
through different ways are close enough using a threshold.  Corruptions can be
detected if the error changes the values such that the difference is greater
than the threshold.  The lower the threshold, the higher the coverage.  

The threshold depends on the rounding error introduced by the ABED tasks (e.g.,
checksum generation) and baseline convolution. 
We explored methods to significantly reduce the error introduced by the ABED tasks. 
Since the filter checksums are generated offline, very high precision
operations can be used to reduce rounding error.
Most architectures support accumulators that use higher precision compared to
inputs (e.g., fp32 accumulation for fp16 input is common). Leveraging such
hardware features, the error in input fmap online checksum generation can be
reduced.  For the FC technique, the resulting checksums are stored with the
filters in fp16 format. The checksums can be stored as multiple fp16 values (or
filters) such that the error introduced by rounding to fp16 is eliminated. For
the FIC technique, the checksum can be stored in a fp32 data type. The output
verification step can also use a higher-precision accumulator to reduce the
rounding error. We estimate the error introduced by the checksum generation and
verification steps (not shown here) and found it to be
much smaller than the average error introduced by the 
convolution. 

Since the error introduced by the baseline convolution is challenging
to bound (due to the varying implementations, algorithms, and input value
distributions), we rely on heuristics to estimate average rounding error. For
some uncommon input values, the observed error can be higher than the threshold
used for error detection, causing check failures in fault-free runs (we call
them false positives).  The system must recover from false positives to
guarantee forward progress.  False positives can be handled using a combination
of techniques, if the rate is low: (1) rerun the layer on a different component
(CPU/GPU/DLA) by incurring higher latency, or (2) notify a higher layer in the
system to determine whether the it can tolerate skipping the inference. A 
recent study found
that many low-level errors are tolerable at the system-level~\cite{Jha2019}.
If the false positive rate is high, a diagnostic module can be invoked to tune
the threshold or switch to the low-throughput duplication mode. 

\section{Conclusions}
\label{sec:conclusions}

CNNs have made their way into safety-critical and HPC systems.
GPU and accelerator-based systems are preferred platforms for CNNs, with
the convolution consuming most of their execution time.
Since safety is paramount for such systems, it is important to ensure the
correctness of convolutions in the presense of hardware errors. 
This paper proposes an algorithm-based error detection
(ABED) solution for convolutions, providing a much lower overhead approach compared to full duplication. We
demonstrate how this solution can be employed during highly optimized
CNN inference executions that fuse multiple layers and use
reduced-precision operations. Results show that ABED 
eliminates convolution output corruptions for all studied hardware errors
with low (6-23\%) runtime overhead, at least 4$\times$ lower than full
duplication.

\bibliographystyle{plain}
\bibliography{bibliography,other_abft}

\end{document}